\documentstyle[12pt]{article}
\newcommand{\ap}{\al^{+i}_{ab}}      \newcommand{\am}{\al^{-i}_{ab}}
\newcommand{\al}{\alpha}             \newcommand{\bt}{\beta}
\newcommand{\ga}{\gamma}             \newcommand{\de}{\delta}
\newcommand{\ep}{\epsilon}           \newcommand{\G}{\Gamma}
\newcommand{\be}{\begin{equation}}
\newcommand{\ee}{\end{equation}}
\newcommand{\bq}{\begin{eqnarray}}
\newcommand{\eq}{\end{eqnarray}}
\begin{document}
\baselineskip= 24 truept
\begin{titlepage}
\title { A New $N=4$ Superconformal  Algebra}
\author{\sc Abbas Ali and  Alok Kumar  \\
\\
 Institute of Physics, Bhubaneswar-751 005, India. \\
 email: abbas, kumar@iopb.ernet.in }
\date{}
\maketitle
\thispagestyle{empty}
\vskip .6in
\begin{abstract}
\vskip .2in
It is shown that the previously known
$N=3$ and $N=4$ superconformal algebras
can be contracted consistently by singular scaling of some of the
generators. For the later case, by a contraction which depends
on the central term, we obtain a new $N=4$ superconformal
algebra which contains an $SU(2)\times {U(1)}^4$ Kac-Moody
subalgebra and has nonzero central extension.
\end{abstract}
\bigskip
\flushright {IP/BBSR/92-90}
\flushright {hep-th/9301010}
\vfil
\end{titlepage}
\eject

It has been known for long time that the contraction \cite{iw} of $SU(2)$
Lie algebra by singular scaling of two of its generators
gives rise to the algebra of $E(2)$. Similar contractions
exist for other groups and are  known as
In\"{o}n\"{u} - Wigner contractions.
It should be emphasized that this contraction
procedure is different from a finite scaling of generators which does
not change the algebra. Due to this characteristic difference, there
are strong restrictions on the choice of the
generators that can be scaled singularly.

In a recent paper \cite{partha}, Majumdar generalized
the contraction procedure to the Kac-Moody algebras
by dividing the generators $J^a$ ($a = 1, \cdots, D$)
into two sets, $J^{\al}$  ($\al = 1, \cdots, d$) and $J^i$, ($i
= d + 1, \cdots, D$)  where $J^i$ are scaled as
\be
	J^i \rightarrow {1\over f(\ep)} J^i\label{scaling}
\ee
with $f(0) = 0,\;\; f(1) = 1$ and $J^{\al}$ are left unscaled. The
contracted algebra corresponds to the limit $\ep \rightarrow 0$.
For the Kac-Moody case, for a consistent contraction to exist, one
gets the condition that the structure constants $f^{\al \bt i} = 0$.
It comes from the fact that a higher order singularity in the right
hand side of a commutator with respect to the one in the
left has to be avoided for a meaningful reinterpretation of
the algebra. The contracted algebra is then given by,
\bq
[J^{\al}_m, J^{\bt}_n] & =&  i f^{{\al}{\bt}{\ga}} J^{\ga}_{m+n}
+ {k\over 2} m {\de}^{\al \bt}{\de}_{m+n, 0}, \\
{}[J^{\al}_m, J^{i}_n] &=&  i f^{{\al} i j} J^j_{m+n},\\
{}[J^i_m, J^j_n] &=&  0.\label{parth}
\eq
It is therefore observed that for the the unscaled
generators, $J^{\al}$, the contraction procedure allows a central
term.  But no such term is allowed for the scaled generators
${J^i}$'s.

In this paper, we analyze the effect of singular scalings of the type
described above for $N=3$ \cite{adem} and Sevrin et al's
$N=4$ \cite{sevrin}
superconformal algebras. The N=3 algebra  contains an $SU(2)$
subalgebra. We find that when the $SU(2)$ generators are scaled as in
\cite{partha}, the consistency of the full algebra requires the
scaling of at least
one of the three superconformal generators. We
present a consistent set of scaling and the corresponding
contracted algebra.

Contraction for the $N=4$ algebra of ref.\cite{sevrin}
is much more  interesting.
In this case we find that the  resulting algebra is
a new $N=4$ superconformal algebra  with an underlying
$SU(2)\times{U(1)}^4$ Kac-Moody subalgebra-with
all the central terms surviving.
For this case, the  scaling is different from the ones used in
previous paragraphs, since it is dependent
on an in-built parameter of the original algebra
which determines the central terms.

To illustrate our basic procedure and for algebraic simplicity,
we first present the contraction of the $N=3$ algebra and later on
go to the more interesting case of $N=4$. We start by writing down
the $N=3$ superconformal algebra:
\bq
[L_m, L_n] & =&  (m-n)L_{m+n}+{c\over {12}}(m^3-m){\de}_{m+n,0}
\label{n3vira},\\
{}[L_m, {\phi}_n] &=& [(d_{\phi}-1)m-n]{\phi}_{m+n},\label{n3wt}
\eq
where ${\phi}_n \in \{G^a_r, J_n^a, \G_r\}$, corresponding $d_{\phi}
\in \{{3\over 2}, 1, {1\over 2}\}$ and $a \in \{1,2,3\}$.
\bq
{}\{G_r^a, G_s^b\} & =& i\ep^{abc}(r - s)J^c_{r+s},
\;\; (a\neq b),\label{n3gg} \\
{}[J^a_m, G^a_r] &=&  m\G_{m+r},\;\; ({\rm no\;sum\;on\;a})\label{n3jg},\\
{}[J^a_m, \G_r] &=&  0 \label{n3jga},\\
{}[J^a_m, J^b_n] &=&  i\ep^{abc}J^c_{m+n}
+ {c\over 3}m\de^{ab}\de_{m+n},\label{n3jajb}\\
{}\{G_r^a, G_s^a\} & =& 2L_{r+s} + {c\over 3}(r^2 -{1\over
4})\de_{r+s, 0}, {\rm (no\;sum\;on\;a)},\label{n3gaga} \\
{}\{\G_r, \G_s\} & =& {c\over 3}\de_{r+s, 0}\label{n3gamgam},\\
{}\{\G_r, G^a_s\} & =& J^a_{r+s},\label{n3gaga0}\\
{}[J^a_m, G_r^b] &=&  i\ep^{abc}G^c_{m+r},\;{(a\neq b)}.\label{n3jagb}
\eq

For the  contraction of this algebra the Kac-Moody generators
$J^a$ are scaled in the same manner as in \cite{partha} with
$J^{\al} \equiv J^3$, $J^i \equiv (J^1,\; J^2)$.
As a result, the modified algebra for these generators becomes,
\bq
[J^3_m, J^3_n] &=& {c\over 3}m\de_{m+n,0}, \label{n3j3j3},\\
{}[J^3_m, J^i_n] &=& i \ep^{ij}J^j_{m+n}\label{jijj},\\
{}[J^i_m, J^j_n] &=& 0. \label{0n31}
\eq
By observing eqns. (\ref{n3gg}) and (\ref{n3gaga0}), we conclude that a
consistent contraction of the $N=3$ algebra then requires the
singular scaling of one of the two pairs of generators,
($G^3$, $\G$) or ($G^1,\;G^2$). We discuss here the former case,
although the later one is also analogous.
The scaling of the four generators $J^1,\; J^2,\; G^3$ and
$\G$ is done by the same factor ${1\over  f(\ep)}$.
Then eqns.(\ref{n3vira})-(\ref{n3jga}) remain
unchanged after scaling. The rest of the algebra,
eqns.(\ref{n3gaga})-(\ref{n3jagb}),
is modified to the following form:
\be
\{G_r^3, G_s^3\} = 0\label{0n32},
\ee
\be
\{\G_r, \G_s\} = 0\label{0n33},
\ee
and
\bq
\{G_r^i, G_s^i\} & =& 2L_{r+s} + {c\over 3}(r^2 -{1\over
4})\de_{r+s, 0}, {\rm (no\;sum\;on\;i)},\nonumber \\
\{\G_r, G^i_s\} &=& J^i_{r+s}, \;\;\; (i= 1, 2)\nonumber,\\
\{\G_r, G_s^3\} & =&  0\nonumber,\\
{}[J^i_m, G_r^3] & =&  0\nonumber,\\
{}[J^i_m, G_r^j] & =& i\ep^{ij}G^3_{m+r},\nonumber\\
{}[J^3_m, G_r^i] &=& i\ep^{ij}G^j_{m+r}.\label{cn3j3gi}
\eq

We have explicitly verified that after the above modifications, all
the Jacobi identities are satisfied. Therefore the modified algebra
(\ref{n3vira})-(\ref{n3jga}), (\ref{n3j3j3})- (\ref{cn3j3gi})
is a consistent superconformal algebra. As in ref.\cite{partha} a
difficulty in the physical interpretation of the above algebra
is the presence
of vanishing (anti-) commutators in eqns.(\ref{0n31}), (\ref{0n32}),
(\ref{0n33}).
It causes  problem for a free field realization of the
algebra and thus for an explicit
construction of the Hilbert space of such theories. Moreover the
vanishing anticommutator (\ref{0n32}) implies that the operator $G^3$ can
not be interpreted as the usual superconformal generator. Similar
problems also occur if one tries to contract $N=1$ and $2$
superconformal algebras. We
now discuss a contraction of the $N=4$ algebra of ref.\cite{sevrin}, where
both of these problems can be avoided.

To work out the contraction of the $N=4$ algebra and to be self contained,
we start by writing down the algebra in \cite{sevrin}. This algebra has
sixteen generators, namely, $L_m$, $G^a_r, \;(a=1,\cdots,4)$,
$A^{\pm i}_m\; (i=1, 2, 3)$, $Q^a_r \;(a=1,\cdots,4)$ and $U_m$. Their
conformal weights $d_{\phi}$ are 2, ${3\over 2}$, 1, $1\over 2$ and 1
respectively. The algebra is written as,
\bq
[L_m, L_n] &=& (m-n)L_{m+n}+{c\over {12}} (m^3-m){\de}_{m+n,0}
\nonumber,\\
{}[L_m, {\phi}_n]&=&[(d_{\phi}-1)m-n]{\phi}_{m+n},
\;\;{\phi}_n \in \{G^a_n, A_n^{\pm}, U_n, Q^a_r\},\nonumber \\
{}[A^{+i}_m, A^{-j}_n]&=&0 \nonumber,\\
{}[A^{+i}_m, Q^a_n]&=&\ap Q^b_{m+n},\nonumber\\
{}[U_m, G^a_n]&=&mQ^a_{m+n},\nonumber\\
{}[U_m, Q^a_n]&=&0\nonumber,\\
{}[U_m, A^{\pm i}_m]&=&0\label{s1},
\eq
and
\bq
\{G^a_m, G^b_n\} &=& 2{\de}^{ab}L_{m+n}+{c\over 3}(m^2-{1\over 4})
{\de}^{ab}{\de}_{m+n,0} 	\nonumber\\
&+& 4(n-m)[ \ga \ap A^{+i}_{m+n} +
 (1-\ga)\am A^{-i}_{m+n}]\nonumber,\\
{}[A^{+i}_m, G^a_n]&=&\ap [G^b_{m+n} -2(1-\ga)m Q^b_{m+n}],\nonumber\\
{}[A^{-i}_m, G^a_n]&=&\am[G^b_{m+n} + 2\ga m Q^b_{m+n}],\nonumber\\
{}[A^{+i}_m, A^{+j}_n]
&=&{\ep}^{ijk}A^{+k}_{m+n}-m{c\over {12\ga}}{\de}^{ij}{\de}_{m+n,0}
\nonumber,\\
{}[A^{-i}_m, A^{-j}_n]
&=& {\ep}^{ijk}A^{-k}_{m+n}
-m{c\over {12(1-\ga)}}{\de}^{ij}{\de}_{m+n,0},\nonumber\\
{}\{Q^a_m, G^b_n\}&=& 2(\ap A^{+i}_{m+n}-\am A^{-i}_{m+n})+{\de}^{ab}U_{m+n}
\nonumber,\\
{}[A^{-i}_m, Q^a_n]&=&\am Q^b_{m+n},\nonumber\\
\{Q^a_m, Q^b_n\}&=&-{c\over {12\ga(1-\ga)}}{\de}^{ab}{\de}_{m+n,0}
\nonumber,\\
{}[U_m, U_n]&=&-m{c\over {12\ga(1-\ga)}}{\de}_{m+n,0},\label{s2}
\eq
with
\bq
\al^{\pm i}_{jk} = {1\over 2} {\ep}_{ijk};\;\;
\al^{\pm i}_{j4} = - \al^{\pm i}_{4j} =
{\pm}{1\over 2}{\de}_{ij};\;\;\al^{\pm i}_{44} = 0,\nonumber
\eq
\bq
{}[\al^{\pm i}, \al^{\pm j}] = -\ep^{ijk}\al^{\pm k},\;\;
{}\{\al^{\pm i}, \al^{\pm j}\}= -{1\over 2}\de^{ij};\;
{}[\al^{+i}, \al^{-j}] = 0.\nonumber
\eq
It is noticed that the central extension of the above algebra is
parameterized by two parameters $c$ and $\ga$. This fact is crucial
for our contraction, since the scaling function $f(\ep)$ for this
case depends on $\ga$. More precisely, we choose $\ep = 1 - \ga >0$,
$f(\ep) = \sqrt \ep$
with $\ga \rightarrow 1$,  and scale eight of the generators
$A^{-i}, \; Q^a, \; U$ of the $N=4$ algebra by the same scaling
function ${1\over {f(\ep)}}$. For $\ga > 1$ one can choose
$\ep = \ga - 1$ and obtain similar results. For our choice,
a close observation  shows
that this is a consistent scaling  for the
algebra in eqns.(\ref{s1}) and (\ref{s2}).
After this singular scaling one gets an algebra where the
(anti-) commutators in eqns.(\ref{s1}) are left
unchanged. The rest of the algebra, eqn.(\ref{s2}), is modified to
\eject
\bq
\{G^a_m, G^b_n\} &=& 2{\de}^{ab}L_{m+n}+{c\over 3}(m^2-{1\over 4})
{\de}^{ab}{\de}_{m+n,0}	\nonumber\\
&+& 4(n-m)\ap A^{+i}_{m+n}\nonumber,\\
{}[A^{+i}_m, G^a_n]&=&\ap G^b_{m+n}, \nonumber\\
{}[A^{-i}_m, G^a_n] &=& 2m\am Q^b_{m+n},\nonumber\\
{}[A^{+i}_m, A^{+j}_n] &=& {\ep}^{ijk}A^{+k}_{m+n}-
m{c\over 12}{\de}^{ij}{\de}_{m+n,0},\nonumber\\
{}[A^{-i}_m, A^{-j}_n]
&=&-m{c\over 12}{\de}^{ij}{\de}_{m+n,0},\nonumber\\
\{Q^a_m, G^b_n\}&=&-2\am A^{-i}_{m+n}+{\de}^{ab}U_{m+n} \nonumber,\\
{}[A^{-i}_m, Q^a_n]&=&0 \nonumber,\\
\{Q^a_m, Q^b_n\}&=&-{c\over 12}{\de}^{ab}{\de}_{m+n,0},\nonumber\\
{}[U_m, U_n]&=&-m{c\over 12}{\de}_{m+n,0}.\label{s3}
\eq

In this case also  we have explicitly verified that the
commutation relations in  eqns. (\ref{s1}) and
(\ref{s3}) satisfy all the Jacobi
identities among these generators. Hence they form a consistent algebra.

	It is noticed that the generators $A^{-i}$'s in the contracted
algebra satisfy the commutation relations of  affine $U(1)$
generators since the  structure constants for these
have disappeared.
This, in fact, is a general property of the contraction
since it was already observed in eqns.(\ref{parth}) that the structure
constants in the commutator of the scaled generators $J^i$'s vanish.
Another  aspect of our  contraction is that
the central terms in all the (anti-) commutators of the original
algebra, eqns.(\ref{s1})-(\ref{s2}) survive and therefore one can
hope to obtain its free field realization.

We have thus obtained a new $N=4$ superconformal algebra which is
distinguished from the previously known $N=4$ algebras of
\cite{adem} and \cite{sevrin}   by the undelying
$SU(2)\times {U(1)}^4$ Kac-Moody symmetry. It is also observed that
the the contracted algebra, eqns.(\ref{s1})
and  (\ref{s3}), contains, as a
subalgebra, the Ademollo et al $N=4$
algebra \cite{adem} which is satisfied by
the generators, $L_m, G^a$, and $A^{+i}$.
Although the existence of Ademollo et
al $N=4$ subalgebra was also noticed in
\cite{sevrin}, however, for their case the choice $\ga=1$
implied $c=0$ and the algebra
became centerless. Here we have shown that
an algebra with nonzero central
extension can be obtained for $\ga \rightarrow 1$
by the singular scaling of eight of
the generators $A^{-i}, Q^a$, and $U$.

Finally, another contraction of the algebra in \cite{sevrin} can also
be done by scaling $A^{+i}$ instead of $A^{-i}$ and taking the limit
$\ga \rightarrow 0$ with $\ep = \ga$.
It will be interesting to obtain a free field realization of the new
$N=4$ superconformal algebra presented in this paper. This is crucial
for obtaining the Hilbert space and the unitary
representations of this algebra.

\vfil
\eject

\vfil
\eject

\begin{thebibliography}{30.}

\newcommand{\npb}{Nucl. Phys. B}
\newcommand{\plb}{Phys. Lett. B}
\newcommand{\prd}{Phy. Rev. D}
\newcommand{\prl}{Phys. Rev. Lett.}
\newcommand{\mpl}{Mod. Phys. Lett. A}
\bibitem{iw} E. In\"{o}n\"{u} and E. Wigner, Proc. Nat. Acad. Sci.(USA)39,
510(1953); R. Gilmore,{\it Lie Groups, Lie Algebras and Some of Their
Applications}, Wiley, New York (1974).
\bibitem{partha} P. Majumdar, In\"{o}n\"{u}-Wigner contraction of Kac-Moody
algebras , Matscience preprint, IMSc/92-26, 1992.
\bibitem{adem} M. Ademollo et al, \plb 62, 105(1976).
\bibitem{sevrin}A. Sevrin, W. Troost, and A. Van Proeyen,
\plb 208, 447(1988).
\end{thebibliography}
\end{document}